\documentclass[12pt]{iopart} 
\usepackage{graphicx}
\def\rtilde{\tilde{r}}
\def\cv{c_{\rm v}}
\def\rhos{\rho_{\rm s}}
\def\Rvir{R_{\rm vir}}
\def\rs{r_{\rm s}}
\def\Deltavir{\Delta_{\rm vir}}
\def\deltac{\delta_{\rm c}}
\begin{document} 
\title[The detection of sub-solar-mass dark matter halos]{The detection of sub-solar 
mass dark matter halos} 
\author{Savvas M. Koushiappas}
\address{Department of Physics, Brown University, Providence, RI 02912, U.S.A.} 
\ead{koushiappas@brown.edu} 
\begin{abstract} 
Dark matter halos of sub-solar mass are the first bound objects to form in cold dark matter theories. 
In this article, I discuss the present understanding of ``microhalos'', their role in structure 
formation, and the implications of their potential presence, in the interpretation of dark matter experiments.  
\end{abstract} 
\submitto{\JPG} 
\maketitle

\section{Introduction}
\label{sec:Introduction}
The term ``microhalo" loosely refers to a generic dark matter, gravitationally-bound object, whose mass is 
typically 
less than that of the Sun.  Objects of sub-solar mass scales are predicted to be the first objects formed in 
the high-redshift Universe in theories where the dark matter particle is ``cold'' (such as $\Lambda$CDM). 
In this short article I summarize the present understanding of microhalos. In Sec.~\ref{sec:decoupling} I 
review the theoretical motivation behind the presence of microhalos, in Sec.~\ref{sec:sims} I present 
the results of numerical simulations that studied their formation. Sec.~\ref{sec:survival} is an overview 
of the present state of understanding the survival of these objects, while in Sec.~\ref{sec:structure} I 
discuss the procedure of  how to characterize their physical properties. I present their connection to direct and 
indirect detection experiments in Secs.~\ref{sec:indirect} \& \ref{sec:direct}, and 
I conclude in Sec.~\ref{sec:conclusions}. 
Throughout the paper, the assumed cosmology is flat $\Lambda$CDM, 
with $\Omega_m h^2 = 0.1358$, $\Omega_bh^2=0.02267$, $h=0.705$, $\sigma_8=0.812$, and a spectral index 
of $n_s=0.96$ \cite{Komatsu:2008hk}.

\section{Kinetic decoupling and microhalo scales}
\label{sec:decoupling}

In the standard cold dark matter cosmological model, the 
energy density in the Universe is balanced  by 
approximately 4\% baryonic matter, 23\% cold dark matter and 73\% dark energy. 
Cold dark matter refers to gravitationally interacting matter that is hypothesized to be in the form of 
a yet-to-be-discovered particle. From the particle physics aspect, new particles arise in essentially 
all extensions to the standard model of particle physics, and usually the lowest mass particle is stable 
due to a new symmetry. 
Dark matter particle candidates include (among many others), neutralinos, the lightest Kaluza-Klein particle, 
axions, and sterile neutrinos. 
Even though there is no {\it a priori} physical reason as to one candidate being more favorable than  
another, neutralinos are attractive and well-studied because they are experimentally accessible at present. 
Neutralinos arise in supersymmetric extensions to the standard model of particle physics 
\cite{Jungman:1995df,BHS05}, and they are part of a generic class of dark matter candidates, called WIMPs 
(Weakly Interacting Massive Particles). 

In the early Universe, conditions are such that a WIMP is in chemical, thermal and kinetic 
equilibrium. As the rates of these interactions are diluted due to the expansion of the Universe, 
the WIMP falls out of equilibrium and decouples. The decoupling temperature is determined by the 
scattering cross section that is responsible for the equilibrium condition the particle decouples from, 
while the velocity distribution function of the particle is set by its mass. 
The temperature at which a WIMP is kinetically decoupled, 
i.e., when momentum-changing interactions cease to be effective, is called the kinetic decoupling temperature. 
For supersymmetric dark matter, such as the neutralino, the kinetic decoupling temperature is $T_{\rm d} 
\sim [10 - 1000]\, {\rm MeV}$ \cite{HSS01,CKZ02,Profumo:2006bv}. The free-streaming of particles after
 kinematic decoupling tends to smooth out fluctuations on small scales 
\cite{Schmid:1998mx,HSS01,Green:2003un,Green:2005fa}. 
This physical effect leads to a cutoff in the dark matter power spectrum. 

\begin{figure}[!t]
\centerline{\includegraphics[width=2.5in]{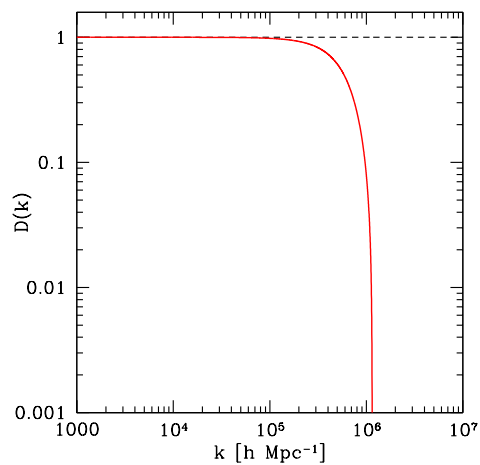}}
\caption{
{\em Left}: An analytic example which shows the damping factor of the power spectrum 
\cite{Green:2003un,Green:2005fa}.}
\label{fig:figure1}
\end{figure}

The cutoff of the power spectrum includes an imprint of the acoustic oscillations of the 
cosmic radiation fluid \cite{LZ05}. In general the transfer function 
of the dark matter density perturbation amplitude is obtained by solving the Boltzman equation, 
but under certain approximations, it is possible to be derived analytically \cite{Green:2003un,Green:2005fa}. 
Fig.~\ref{fig:figure1} shows an example of the analytic calculation of the cutoff of the 
power spectrum, derived from \cite{Green:2003un,Green:2005fa}. What is shown is the damping factor $D(k)$ that arises for a supersymmetric WIMP 
dark matter particle with a 
decoupling scale of $k_{\rm d} \sim 40 \, {\rm pc}^{-1}$, and a 
free-streaming scale of $k_{\rm fs} \sim 0.94 \, {\rm pc}^{-1}$ (corresponding to a kinetic decoupling 
temperature of $T_{\rm d} \sim 20 \, {\rm MeV})$.  
In general, the cutoff scale is related to the kinetic decoupling temperature as 
$M_{\rm cut} \approx 10^{-4} M_\odot (T_{\rm d} / 10 \, {\rm MeV})^{-3}$ \cite{LZ05}. 
Recently, \cite{Martinez:2009jh} presented the posterior of the cutoff scale from a 
Bayesian analysis of the presently-allowed supersymmetric parameter space. 

\section{Microhalos in numerical simulations}
\label{sec:sims}

\subsection{Formation of microhalos}

The first attempt to numerically simulate the formation of microhalos was done in 2005 by the group at 
the University of Z\"{u}rich \cite{DMS05}. In that work, a small region of the Universe was simulated 
using a multi-scale technique \cite{Bertschinger:2001ng} in order to achieve the required spatial resolution 
needed to resolve objects below the cutoff-scale of the power spectrum. The input transfer function was 
taken from \cite{Green:2003un}; corresponding to a supersymmetric dark matter candidate of mass $\sim 100 
{\rm GeV}$. 
This particular choice leads to an exponential cut-off at $\sim 10^{-6} M_\odot$. The simulation was 
evolved from a redshift of $z=350$ to a redshift of $z=26$, when the simulated region itself approached   
non-linearity. 
In the small, high-resolution volume of roughly average density, the first non-linear structures with mass 
near the cutoff scale of $\sim 10^{-6} M_\odot$ are observed to form by redshift $z \approx 60$. 
Their physical size is few $\times 10^{-2}$ pc, 
and they are described by a single power-law density profile. Even though halos of mass in the range $[10^{-6} - 
10^{0}] M_\odot$ are formed roughly at the same time, they all share the same concentrations at $z=26$. 
As expected, these objects contained no substructure, and no objects with smaller mass were formed, confirming 
the input shape of the power spectrum. Even though the statistics in this first simulation were low, the 
deduced mass function suggested a power law and normalization consistent with the halo mass function derived 
from numerous orders of magnitude larger scales, namely, $dN/d \ln M \sim M^{-1}$. This finding is an important 
piece of information when it comes to studies aimed at the long-term survival of microhalos in the potential 
well of a present-day galaxy, such as the Milky Way. 

In a subsequent paper \cite{Diemand:2006ey}, the numerical simulation of the formation and evolution of 
microhalos was studied by using 64 million dark matter particles within a comoving volume box of 3 kpc on the side  
(implying a particle mass of $9.8 \times 10^{-10} M_\odot$). The initial redshift was $z=456$, and the 
initial conditions were similar to the simulation discussed in \cite{DMS05} (see previous paragraph). 
By the end redshift of the simulation, $z\approx 75$, the volume contained almost 2000 virialized dark matter
sub-solar mass halos. The most massive halo is $\sim 10^{-2} M_\odot$, and corresponds to a 3.5$\sigma$ 
fluctuation in the density field. For a comparison of the evolution and substructure content of this 
simulation, the authors performed a separate simulation, of similar dynamic range, of a cluster-size 
halo at $z=0$. Comparison between the substructure mass functions in the two simulations showed confirmation 
of the earlier result found in \cite{DMS05}, namely a power law behavior that is self-similar over more 
than 10 orders of magnitude in subhalo mass. 

It should be emphasized that the simulation presented in 
\cite{DMS05} was performed in a region of average density, thus probing the formation of ``field'' 
sub-solar mass dark matter halos. 
On the other hand, the high-resolution 
simulation presented in \cite{Diemand:2006ey} was centered around the highest density peak of the 
initial 3 kpc volume. This allowed the study of the assembly and structure of an object with mass 
higher than the cutoff scale of the power spectrum, while resolving scales down to the cutoff scale. 
In addition, it provided the first estimate of the survival rate of dark matter microhalos, a very important 
result which relies on the non-linear evolution of structure formation.

\subsection{Challenges in numerical simulations}
The numerical simulation of the formation and survival of microhalos is a very difficult task. 
The difficulties arise from two main sources. First, by the fact that the power spectrum $P(k)$ of sub-solar 
mass scales asymptotically approaches 
the $k^{-3}$ behavior, thus leading to the formation of multiple scales roughly at similar epochs, and  
second, by uncertainties in the actual implementation of numerical techniques in numerical simulations 
\cite{Elahi:2008cd}. 

The cutoff in the power spectrum that enters the initial conditions in the 
numerical simulations does not fully suppress the formation of structure at scales smaller than the 
cutoff scale. Even though the fraction of substructure in objects below the cutoff scale is small, their 
mere presence deserves further study in order to determine whether this effect is an artifact of the 
simulation techniques employed, or whether it is a physical effect. It should be noted that such effects 
have also been found in simulations of Warm Dark Matter (WDM), where the dark matter particle possesses a 
non-zero thermal velocity \cite{Bode:2000gq,Knebe:2003hs,Yoshida:2003rm,Colin:2000dn}. 
A possible numerical source that leads to this artificial effect is the implementation of  
grid initial conditions \cite{Goetz:2002vm}. If not implemented at an early enough starting redshift, 
the initial Zel'dovich displacement can be a large fraction of the inter-particle separation 
\cite{Heitmann:2006eu}. This can lead to artificial fragmentation along filaments, 
however the effect has also been observed when 
the use of glass initial conditions \cite{1996clss.conf..349W} is implemented. The 
simulations of  \cite{Diemand:2006ey} 
also show that the formation of sub-cutoff scale structure is not limited to a preferred separation along 
a grid axis, suggesting that it may perhaps be a real effect. 

Another source that can introduce bias in the 
measurement of a mass function is the starting redshift of simulations. As shown in \cite{Heitmann:2006eu}, 
the requirement that the Fourier modes within the simulation box are all linear is very naive. The mass 
function cannot be robustly determined at redshifts which are close to the redshift of first crossing (the 
redshift where halos form). 
It has also been pointed out that even perhaps the choice of a halo finder 
in the simulation can also lead to an uncertainty in the measurement of the mass function 
\cite{Diemand:2006ey,Elahi:2008cd}, due to the fact that the density contrast in sub-solar mass halos is low due 
to similar formation times. 
Therefore, caution must be taken in the interpretation of these results, and a thorough convergence 
study on the mass function of sub-solar mass halos is of paramount importance, and in great need.

\section{Survival of microhalos in the Milky Way halo} 
\label{sec:survival}

The survival (and thus existence) of microhalos in the present-day Milky Way halo has been a subject of debate. 
The approach typically taken in these studies has progressed in two fronts: simple analytical estimates, and 
through numerical simulations. In principle the two should agree, however given the vast orders 
of magnitude in dynamic range and highly complex non-linear effects, it has proven to be a difficult task. 

The question of surviving microhalos was first addressed analytically in \cite{Berezinsky:2003vn}. 
A detailed analytical study of the effects of tidal destruction on microhalos due to rapid, 
early-time hierarchical merger phase, showed that only $[0.1-0.5]\%$ of formed microhalos would survive 
to the present epoch \cite{Berezinsky:2003vn}. The initial claim 
that a large number of microhalos could 
be present in the solar neighborhood \cite{DMS05} 
was challenged only days later \cite{ZSH1}. The argument presented was that strong 
impulses by individual stars in the disc can tidally disrupt microhalos, leading to a large number of 
tidal streams. Within days, a response \cite{METAL05} showed that the application of the impulse 
approximation requires microhalos to experience multiple stellar interactions and as a result the survival 
timescale is many Hubble times. 

One of the key advantages of numerical simulations in the study of the formation and abundance of 
microhalos is the ability to trace the history of each bound structure. 
In doing so, the work of 
\cite{Diemand:2006ey} found that substructure formed at early times is highly prone to tidal disruption 
due to the high central density of the host halo. In addition, most of the remaining microhalos that survived 
to the end-redshift of the simulation run are significantly tidally stripped. As expected, the amount of 
mass-loss experienced due to tidal effects depends on the history of the subhalo and in particular whether 
it experienced passages near the center of the host halo. However, most importantly, the study of 
\cite{Diemand:2006ey} demonstrated that even though the density contrast of sub-solar mass scales is small, 
{\em microhalos of mass $10^{-6} M_\odot$ are able to survive in a host of mass at least $10^4$ times more massive}, 
thus addressing numerically the question of survival in the initial merging process.

A separate study \cite{Green:2006hh} looked at the  
energy input to microhalos due to stellar interactions by using analytic calculations, and 
then tested their results in a numerical simulation \cite{Goodwin:2003fn,Goodwin:2004yb}. 
The main result is that at large impact parameters, the energy input to microhalos is independent of mass. 
They show that the survival timescale of microhalos in the Milky Way is of order the age of the Milky Way, 
but decreases significantly as the microhalo mass increases. In an independent test of the impulse 
approximation \cite{Goerdt:2006hp}, it was shown 
that multiple stellar encounters can remove most of the mass from microhalos, however the very inner cores 
can survive to the present time. Nevertheless, the normalization and shape of the density distribution of 
dark matter in microhalos will be decreased, with significant implications to direct and indirect detection 
experiments (see following sections). It is important to realize that it is currently not feasible to fully address the 
issue of survival, mass and spatial distribution of microhalos, due to the enormous dynamic range required and 
the limited computing power that technologically exists. Studies such as the ones mentioned here are 
invaluable in shedding light on what might happen in ideal situations, and therefore should be further explored.

\section{Characterizing the physical properties of microhalos}
\label{sec:structure}

In order to assess the detectability of microhalos in any scheme (e.g. direct or indirect detection experiments), 
it is first necessary to 
determine the distribution of dark matter within the virialized region that comprises the 
microhalo. Typically, this can be done by assuming a profile, and a normalization of that profile. 

The numerical work of \cite{DMS05,Diemand:2006ey} finds that microhalos {\it at the redshift at which 
they were studied in the simulation} are described by an NFW profile \cite{NFW96}. The form of this profile 
is $\rho(r) = \rhos / \rtilde ( 1 + \rtilde)^2$, where $\rhos$ is a density normalization, called the 
characteristic density, $\rs$ is the scale radius, and $\rtilde = r / \rs$. The two parameters needed to 
specify the profile are the characteristic density, and the scale radius. The numerical simulation 
findings suggest that $\rs$ is approximately equal to the virial radius $\Rvir$ of the microhalo, suggesting 
an NFW concentration parameter ($\cv = \Rvir / \rs$) {\it at that redshift} which is of order one (more 
specifically $\cv \approx [2-4.5]$ \cite{DiemandPC}). For field microhalos, the concentration parameter grows 
as $(1+z)^{-1}$, indicating that ``field'' subhalos have concentrations in the range of $[50-100]$ for a formation 
redshift around $z \sim 20$ (see below). 
Obviously, the profile of a 
present-day microhalo in the Milky Way has most likely evolved from this initial distribution of dark matter, 
to a distribution which is the result of tidal interactions throughout the course of structure formation. 
As I alluded to in Secs~\ref{sec:sims} \& \ref{sec:survival}, 
these processes are very difficult to accurately model. Nevertheless, 
it is a reasonable assumption to consider the initial profile at the redshift of formation 
as a representation of a maximal distribution, 
and keep in mind that any experimental ramifications are going to actually be upper limits and not 
absolute values. 

In order to normalize the dark matter density profile of microhalos we need to first determine their 
{\it formation time}, which can be obtained by knowledge of the CDM power spectrum, and the 
cosmological growth factor.

The normalization of a the profile of a dark matter halo can be obtained by assuming that 
mean density of the halo is $\Deltavir$ times the mean matter density of Universe $\rho_{\rm M}$, 
with $\Deltavir$ being the 
virial overdensity (see e.g. \cite{Bryan:1997dn}). The mass of the halo is then 
$M = 4 \pi \rho_{\rm M} \Deltavir(z) \Rvir^3/3$. By using this definition of the mass of a dark matter halo, the 
normalization of the NFW profile is then simply obtained via $\rhos = \rho_{\rm M} \Deltavir \cv^3 / 3 f(\cv)$, 
where $f(x) \equiv \ln (1 + x ) - x / ( 1 + x )$. 
\begin{figure}
\centerline{\includegraphics[height=2.5in]{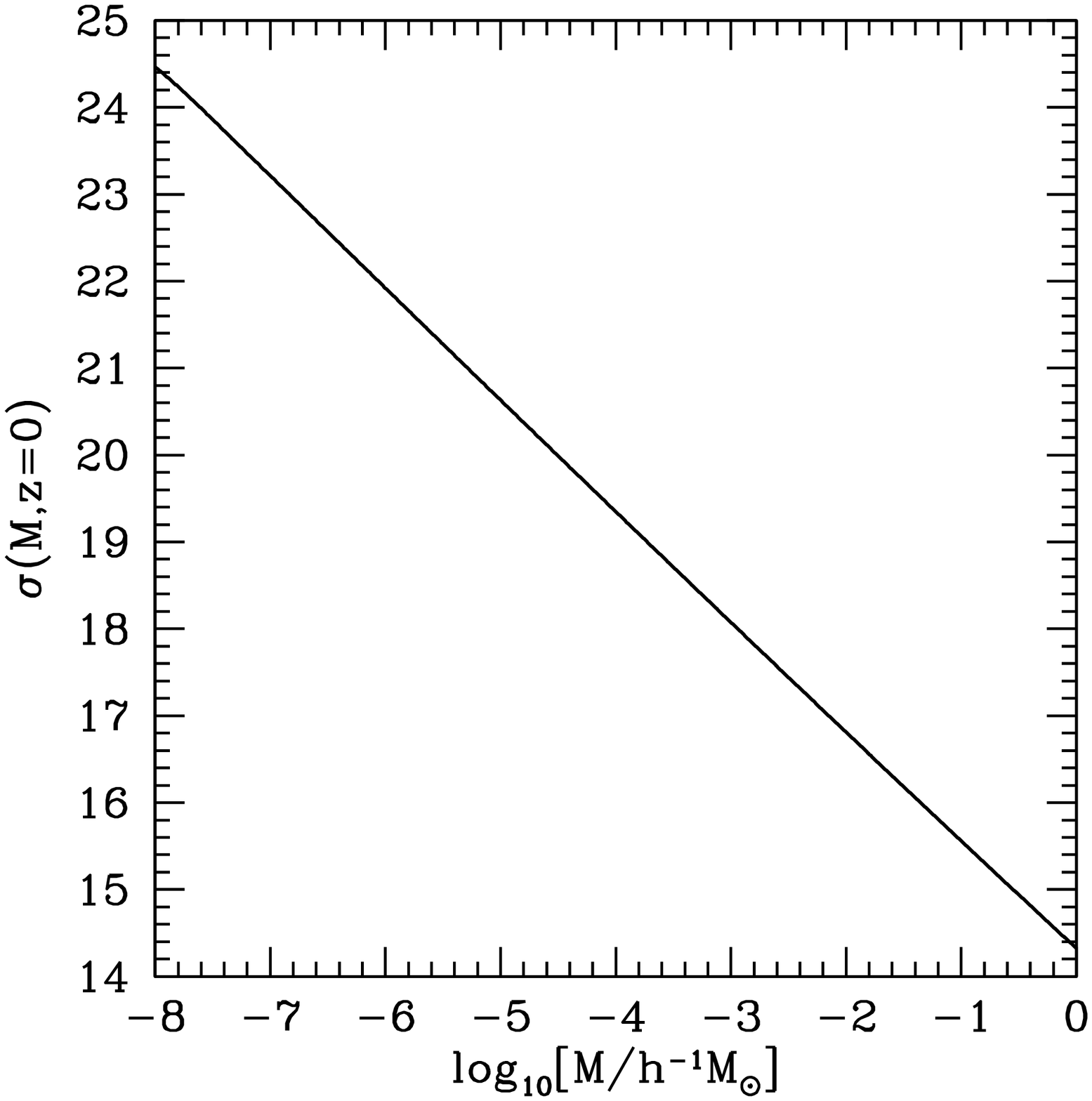}
\includegraphics[height=2.5in]{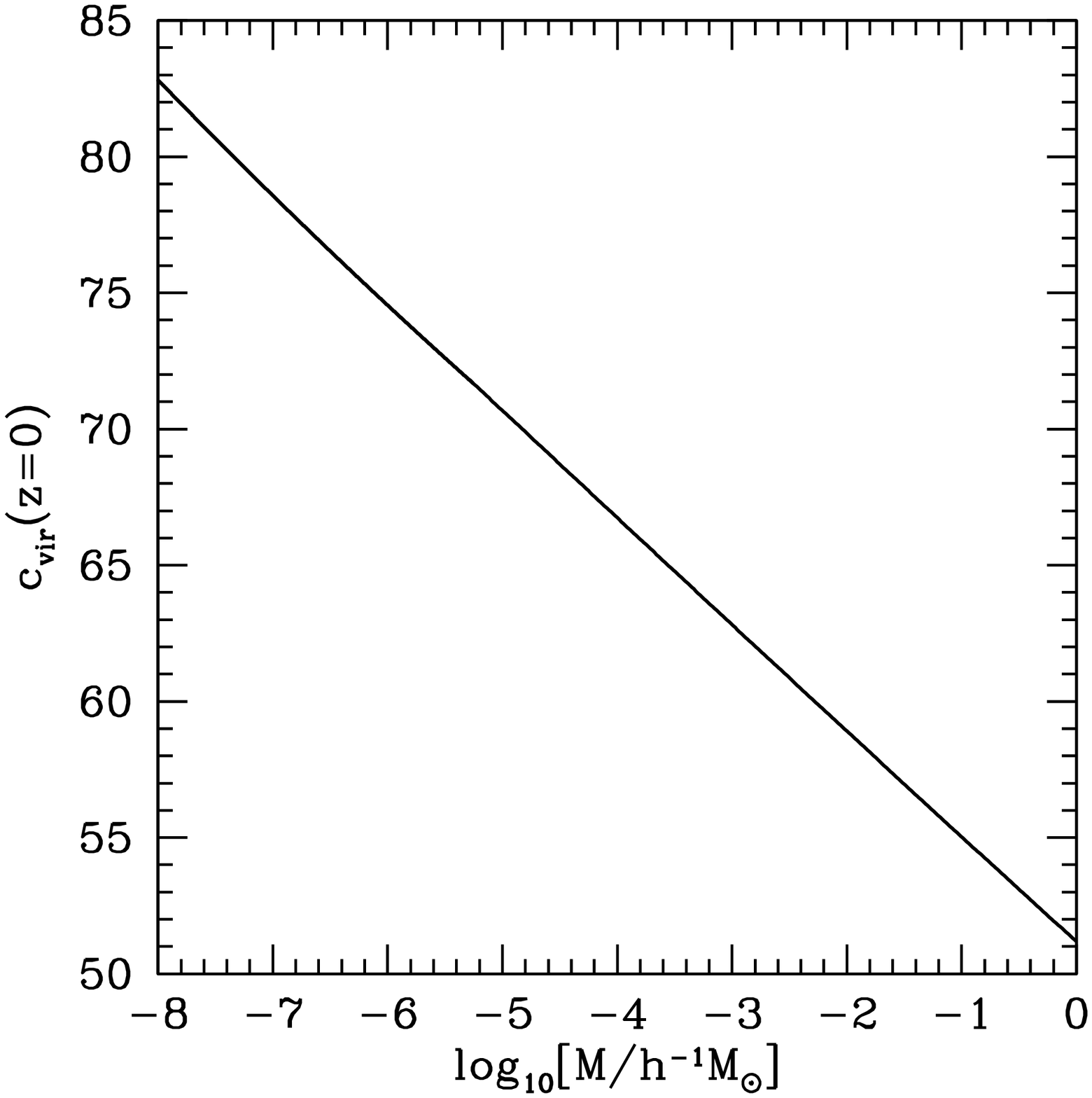}}
\caption{ 
{\em Left}: The rms fluctuation on sub-solar mass scales for the WMAP5 cosmology.  
{\em Right}:  The concentration parameter of ``field'' sub-solar mass halos at $z=0$ based on the 
model of \cite{Bullock:1999he}. This is obtained from an extrapolation of   
the $\cv(M)$ relationship found on galactic scales, all the way down to microhalo scales. 
It does not represent the $\cv(M)$ relationship of Milky Way progenitors that were of microhalo scale at the time of their formation.}
\label{fig:figure2}
\end{figure}
The rarity of a particular mass scale at a particular redshift  can be obtained from the power spectrum of 
fluctuations. In general, if we define the power in each logarithmic interval in $k$ as $\Delta^2(k) \sim 
k^3 P(k)$, where $P(k)$ is the CDM power spectrum, 
then the mean square 
fluctuations on a scale $M$ is given by $\sigma^2(M) = \int \Delta^2 (k) | W(k;M)|^2 d \ln k$. Here, 
$W(k;M)$ is the Fourier transform of a real-space spherical top-hat window that contains mass $M$. 
The left panel in Figure~\ref{fig:figure2} shows the root mean square fluctuations on microhalo scales, derived by 
assuming the WMAP5 cosmological parameters \cite{Komatsu:2008hk}. Note that for illustration purposes, 
the primordial power spectrum in this example does not contain the cutoff shown in Fig.~\ref{fig:figure1}. 
Knowledge of the underlying cosmological model (i.e., the values of the cosmological parameters), allows us 
to compute the growth function of fluctuations $D(z)$. 
Then, the redshift of formation of a mass scale $M$ 
is defined as the redshift $z$ where $\sigma(M)D(z) = \deltac$, where $\deltac = 1.686$ is the value of the 
characteristic density for collapse (derived from linear theory).  The rarity of a particular peak is typically 
defined to be in units of the standard deviation of the smoothed density distribution, so that 
$\nu = \deltac / \sigma(M) D(z)$. 

Note that the {\it rarity} of a particular fluctuation in the density field is directly related to 
the {\it abundance} of that scale in the present Milky Way halo - a key quantity in any dark matter 
experimental effort. 
For example, a smaller fraction of the total Milky Way mass originates from 3$\sigma$ peaks in the density field as 
those are rarer than e.g. $1 \sigma$ peaks. However, the 3$\sigma$ peaks are formed earlier, thus they will 
be denser. As such, they 
are not as susceptiple to tidal disruption as the $1 \sigma$ peaks. The abundance of microhalos in the present 
epoch will be determined by a balance of these competing effects and a full description of the microhalo population 
of the Milky Way {\em must} include all of these factors. A first step in that direction was the study of 
\cite{Diemand:2005rd}, where it was found that high-$\nu$ progenitors are found predominantly found near 
the centers of dark matter halos today. For example, according to \cite{Diemand:2005rd}, the median 
progenitor at the solar neighborhood of the Milky Way corresponds to a $2\sigma$ peak.

\begin{figure}[!t]
\centerline{\includegraphics[width=2.5in]{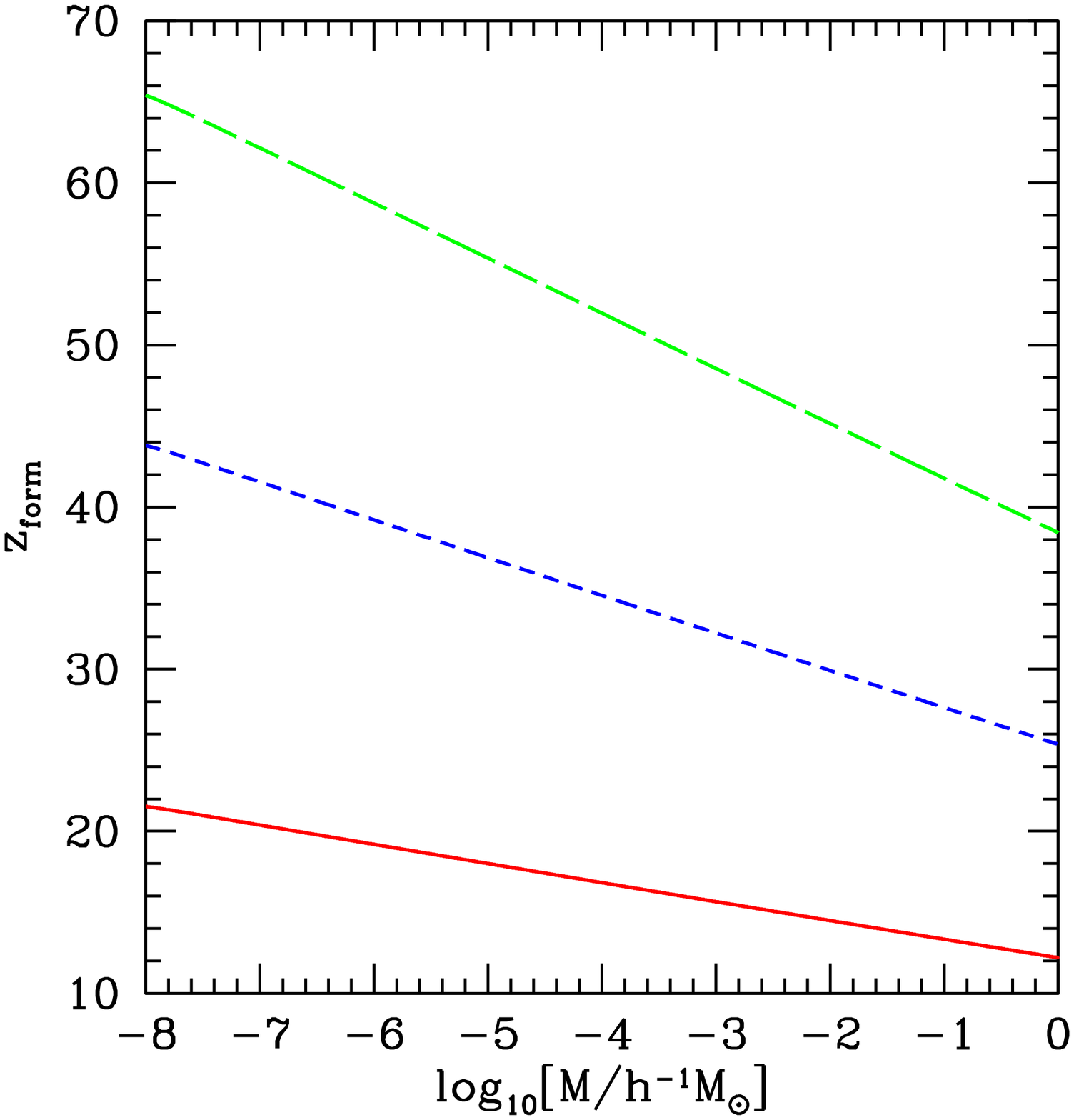}
\includegraphics[width=2.5in]{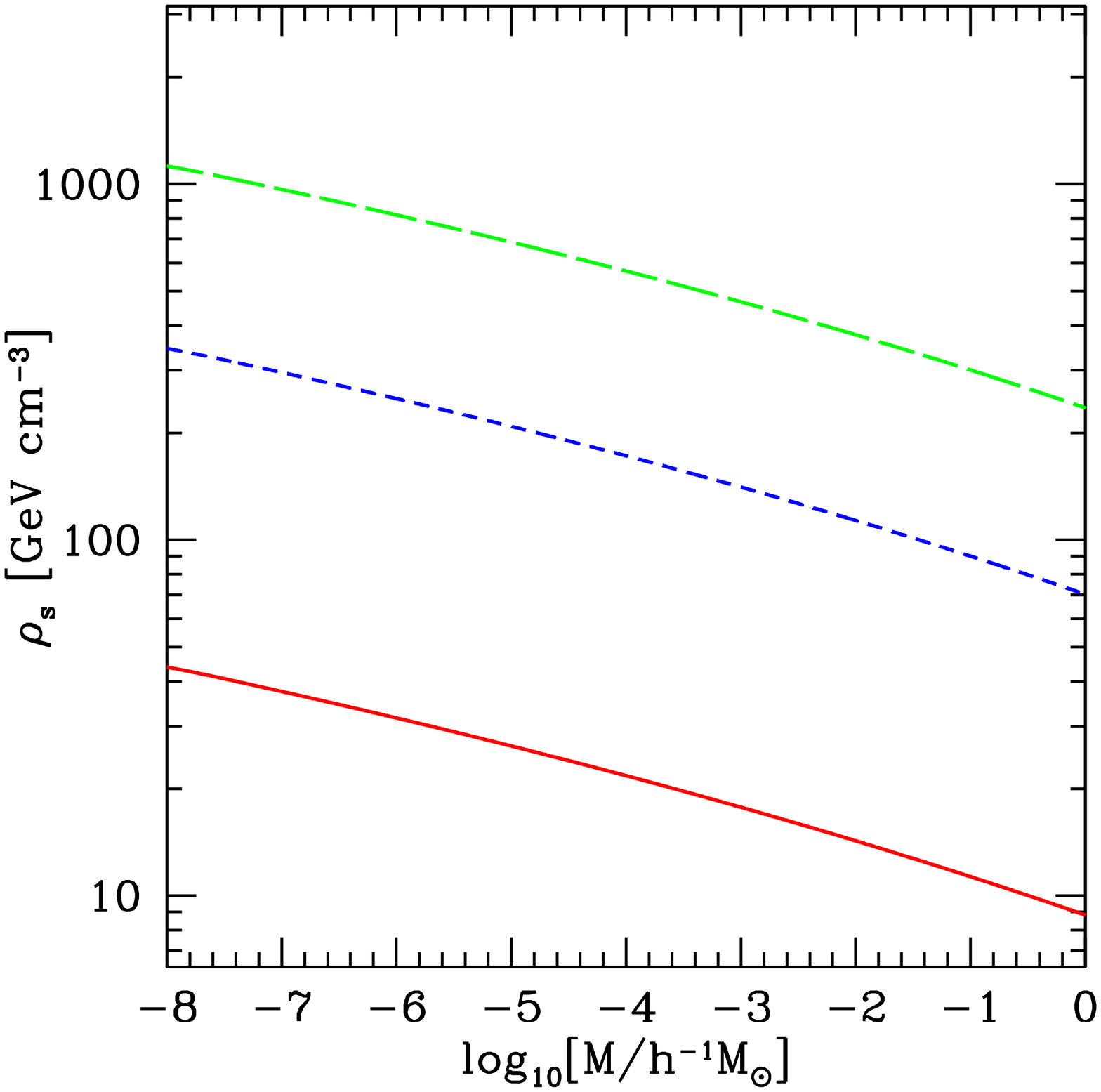}}
\caption{ 
{\em Left }:  The formation redshift (according to linear theory) as a function of halo mass. The curves correspond 
to 1 (solid red), 2 (short-dash blue) and 3 (long-dash green) $\sigma$ fluctuations in the density field.
{\em Right }:  The NFW characteristic density of halos collapsing out of 1, 2, and 3 $\sigma$ peaks. 
In all cases, the underlying cosmological model is taken from the 
combined analysis of WMAP5 with distance measurements from the Type Ia supernovae, and the Baryon Acoustic 
Oscillations \cite{Komatsu:2008hk}.}
\label{fig:figure3}
\end{figure}

An additional piece of information which is of importance when it comes to dark matter detection experiments, 
is the 
tidal radii of survived microhalos in the Milky Way. A reasonable approach 
would be to assume that the truncation radius at 
the present epoch can be estimated by finding the radius of the microhalo where the density is equal to the 
density at the solar radius \cite{Klypin:2001xu} . This assumption results in the truncation radius being 
equal to the virial radius 
for the microhalos halos considered here, reflecting the dense state of the Universe at the time they were formed.
Clearly, the assumption here is that the halo profile normalization 
is intact from the time it was 
formed until the present epoch. Again, this points out at the lack of knowledge of the mass function and history (thus 
properties) of microhalos in the solar neighborhood. But given the high initial densities reflected in microhalos, and 
the steep power law of the density profile,  it seems likely that 
the tidal radius will not be very different from their virial radius set at the time of collapse.

A more general note regarding the physical description of the dark matter profiles of microhalos is with regards to 
the concentration parameter. 
It is dangerous to assign concentration 
parameters that would correspond to present-epoch ``field halos'' (i.e., a microhalo formed at high 
redshift and evolved uninterrupted to the present day). Field halos grow hierarchically by the accretion of 
smaller halos, and by smooth accretion of dark matter.  
It is not clear exactly how this process occurs at very high redshifts on microhalo scales, 
as the flatness of the 
power spectrum implies the simultaneous collapse of many different scales above the cutoff in the power spectrum. 
The {\it maximum value} of the concentration parameter for a field microhalo can be obtained by extrapolating 
the concentration-mass relationship deduced from numerical simulations \cite{Bullock:1999he,Eke:2000av} (see 
Fig.~\ref{fig:figure2}). Even though 
this is a daring extrapolation over numerous orders of magnitude, it can be considered as an upper limit for 
{\it field microhalos}, as the 
concentration-mass relationship stems from the shape of the power spectrum. The power spectrum flattens out at 
large $k$, and so is $\cv$ with respect to mass \cite{Bullock:1999he} (see Fig.~\ref{fig:figure2}). 
The danger from assuming concentration parameters of field halos in microhalo studies near the solar 
neighborhood arises due to the fact that the density 
profile normalization is rather sensitive to the value of the concentration parameter $\rhos \sim \cv^3$, 
therefore 
assuming concentration parameters of order 100, can result in a very large overestimate of the normalization 
of the profile.  

\begin{figure}[!t]
\centerline{\includegraphics[width=2.5in]{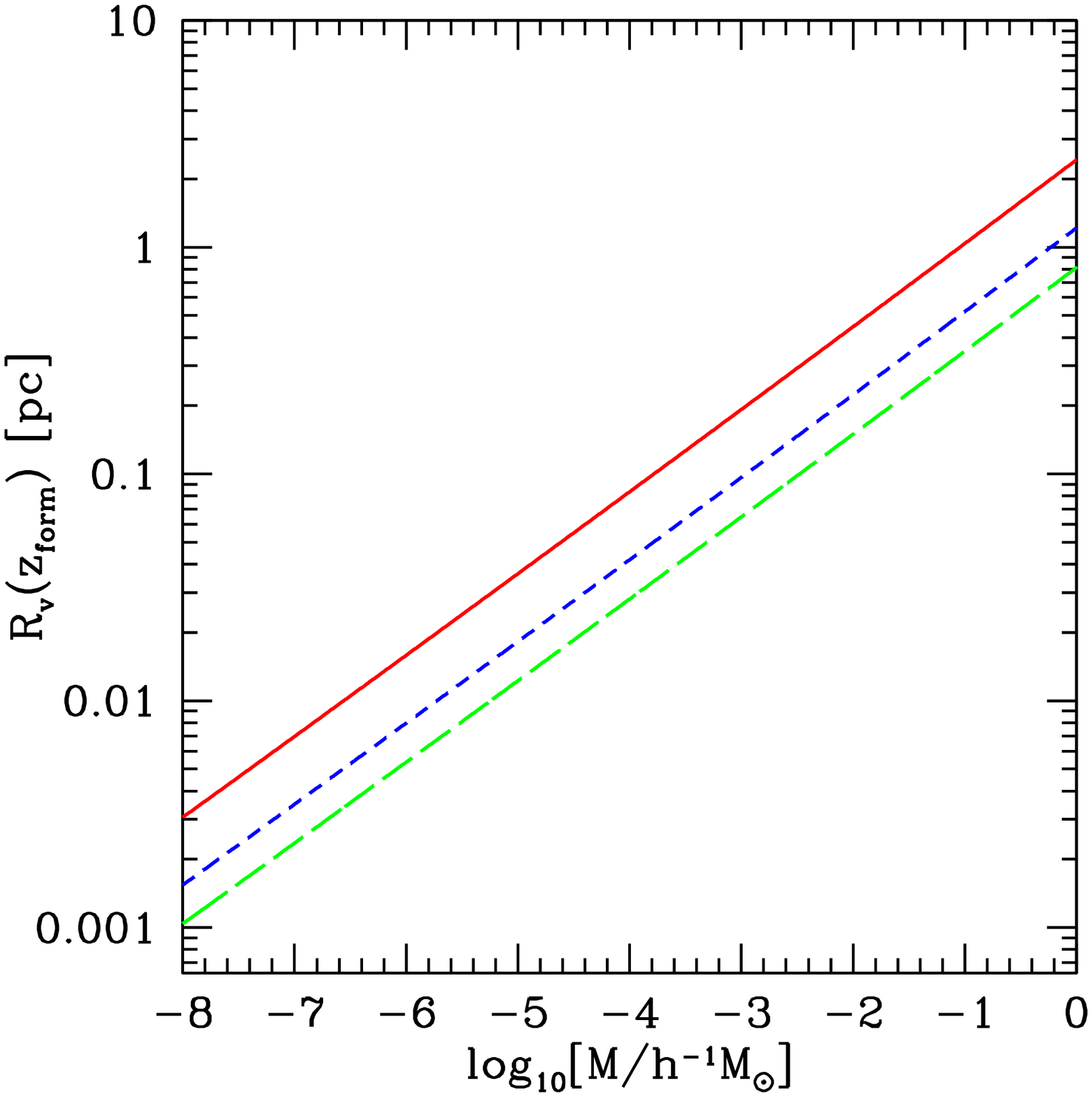}
\includegraphics[width=2.5in]{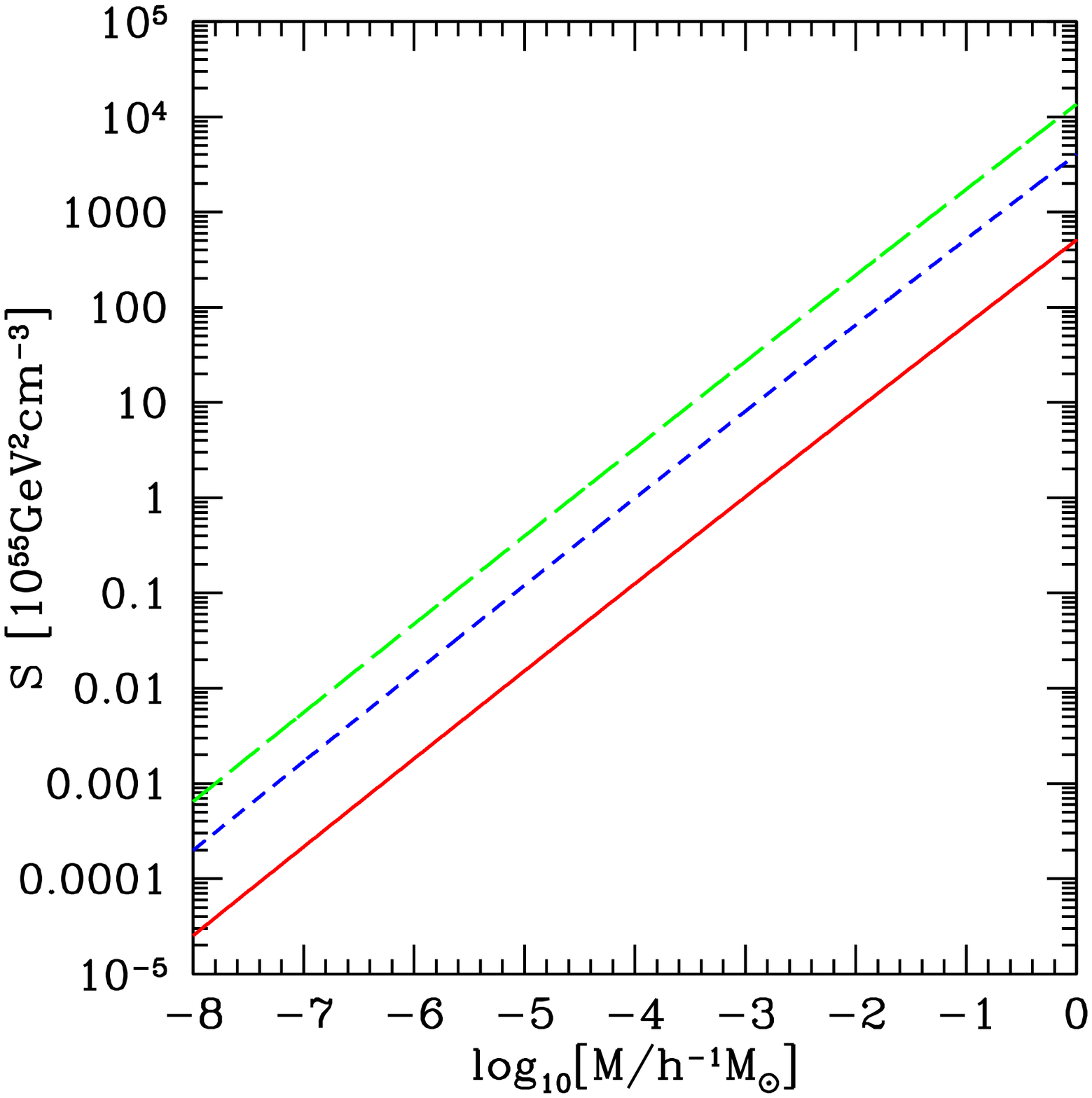}}
\caption{{\em Left}: The virial radius of microhalos at the redshift of collapse.
{\em Right}: The annihilation luminosity of a sub-solar mass dark matter halo as a function of its mass. 
If the halo has lost 99\% of its mass due to tidal interactions with other halos and/or baryons (stars, disk, 
clouds, etc.), then the luminosity should be decreased by a factor of 30\% from the values shown. Line 
types and colors as in Fig~\ref{fig:figure2}.}
\label{fig:figure4}
\end{figure}

To summarize, given the uncertainty in the evolution of the profile of microhalos, 
the only reasonable approach is to determine their physical properties at the time they were formed, 
and consider those as upper limits to what they could be at the present epoch. 
Figure~\ref{fig:figure3} \& \ref{fig:figure4} 
shows the formation redshift, characteristic density and virial radius of sub-solar mass 
dark matter halos. It is of vital importance to emphasize that {\it any} treatment of 
annihilation signals 
from microhalos (see e.g. $\gamma$-rays in the next section) must be consistent with a description of the 
physical properties of microhalos, which are set by the underlying structure formation paradigm. 
Arbitrary choices of concentration parameters (which affect the normalization 
of the profile), or inconsistent choices of formation redshift will only result in an estimate for a 
particular toy microhalo model. 
Such an approach cannot assess the structure formation relevance, and/or distribution function of expected 
signals and must treated with caution.

\section{Detection of microhalos using dark matter annihilation products}
\label{sec:indirect}

It might be possible to detect the presence of microhalos in the Milky Way halo by searching for the annihilation 
products of the WIMP dark matter particle, and specifically for $\gamma$-rays. 
Under the supersymmetric WIMP scenario, the dark matter particle couples to quarks and leptons at the 
tree level, and to photons and the $Z^0$ gauge boson via one-loop diagrams. 
The search of $\gamma$-rays from the annihilation of the dark matter particle is attractive for two reasons: 
1) a swath of $\gamma$-ray experiments are operational (e.g. Fermi/GLAST \cite{2005AAS...207.2405R}, 
VERITAS \cite{Weekes:2001pd}, H.E.S.S. \cite{2003ICRC....5.2811H}, MAGIC \cite{2003ICRC....5.2815M}, 
CANGAROO \cite{Yoshikoshi:1999rg}), and 
2) the shape of the emitted spectrum can be estimated (given a supersymmetric dark matter candidate). 

In 
general, the luminosity of a subhalo as a function of energy is given by $L(E) = S \, \int P(E) \,dE$, 
where $S = \int \rho^2(r) d^3r$ 
(see Fig.~\ref{fig:figure4}), and 
$P(E) = (dN/dE) \langle \sigma v \rangle / M_\chi^2$. Here,  $dN/dE$ is the energy spectrum 
of the annihilation products under 
study (e.g. photons, positrons, neutrinos, etc.),  
$\langle \sigma v \rangle$ is the annihilation cross 
section, and $M_\chi$ is the mass of the dark matter particle. Maximal values of $\int P(E) dE$ in supersymmetric 
theories with a valid dark matter candidate (e.g. the neutralino) yield a value of $\le 10^{-28}
\, {\rm cm}^3 {\rm GeV}^{-2}$ for a threshold energy of $\sim 1$ GeV.

\subsection{Microhalos as $\gamma$-ray sources} 

\begin{figure}[!t]
\centerline{
\includegraphics[width=2.5in]{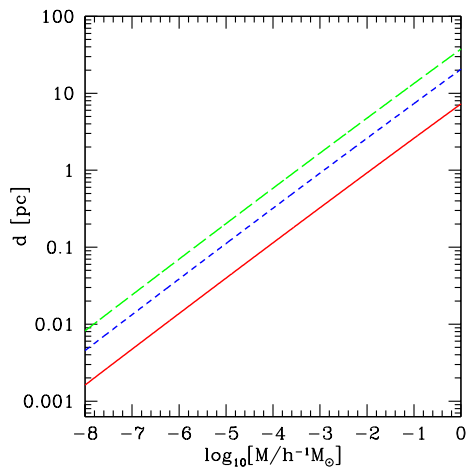}}
\caption{The visibility distance where the emitted flux is greater than the threshold flux of Fermi/GLAST 
as a function of microhalo mass for different peaks of the density 
field (see text for details). Line types and colors as in Fig.~\ref{fig:figure2}.}
\label{fig:figure5}
\end{figure}

In this subjection I discuss the possible detection of microhalos with $\gamma$-rays, though the formalism 
can be well-extended to other annihilation products, such as positrons and neutrinos. 
In order for a sub-solar mass halo at a distance $d$ and flux on Earth $\Phi = L / 4 \pi d^2$ 
to be detected above a detector threshold $\Phi_0$, it must be located at a
distance $d< \sqrt{ \Phi/ \Phi_0}$. Such an example is shown in Fig.~\ref{fig:figure5}, for a supersymmetric WIMP 
dark matter candidate with   $P \le 10^{-28}\, {\rm cm}^3 {\rm GeV}^{-2}$, and a detector threshold of 
$\Phi_0 = 10^{-9} {\rm cm}^{-2} \, {\rm s}^{-1}$. 
It should be noted that for subhalo mass functions that scale as $dN/d \ln M dV \sim M^{-1}$, the typical 
visibility distance between microhalos of mass $M$ scales as $\langle r \rangle \sim M^{1/3}$. 
However, as the luminosity 
of a halo goes as $L \sim \rhos^2 \rs^3 \sim M$, the ``visibility'' distance of microhalos is 
proportional to $d_{\rm max} \sim M^{-1/2}$, i.e., it is ``easier'' to detect more massive halos at larger 
distance \cite{KZW04,Kuhlen:2008aw}, than small halos nearby. 
This scalings will be altered if the dark matter particle's annihilation 
cross section is Sommerfeld enhanced 
\cite{Sommerfeld:1931,Hisano:2003ec,Profumo:2005xd,ArkaniHamed:2008qn,MarchRussell:2008tu}. 
The increased annihilation rate due to low microhalo internal velocities 
leads to the possibility that small nearby microhalos will be more luminous than the canonical case where the 
annihilation cross section is independent of velocity. 

One approach to investigate the observability of a population of 
microhalos in the Milky Way is to Monte Carlo the distribution and properties of 
microhalos as followed in \cite{Pieri:2005pg}. The key unknown in this approach (or any approach for that 
matter) is the abundance of microhalos as a function of their mass, as well as their central densities (as there is a 
a distribution of mass scales that collapse in each redshift interval - see Sec.~\ref{sec:structure}). 
The results of \cite{Pieri:2005pg} show that for a fixed concentration microhalos (i.e., isolated 
density peaks), it is highly unlikely 
that there should be any high signal-to-nose single-source detection of microhalos 
with the all-sky Fermi Gamma-ray Space Telescope (FGST, formerly GLAST). 

Another approach was followed in \cite{Koushiappas:2006qq}. In that study, the uncertainty that stems from the lack 
of knowledge of the survival rate of microhalos was parametrized as a fixed contribution to the local 
dark matter density, so that a particular fluctuation peak (i.e., collapsed microhalo at a particular redshift) 
results in a certain abundance. Again, the main caveat here is the unknown survival rate. 
The result of \cite{Koushiappas:2006qq} suggested that a large number of microhalos could potentially be 
visible, albeit the normalization of the number was directly deduced from the unknown density parametrization. 

One unique feature of the potential presence of survived microhalos in the Milky Way is the probability 
that they can be present in very short distances (sub-parsec) from the Solar system. Thus, 
it may be possible to observe a proper motion of the $\gamma$-ray signal \cite{METAL05,Koushiappas:2006qq,
Koushiappas:2007mx}. The two crucial conditions for the observation of the proper motion of a subhalo is that 
it is bright enough to be detected at some high signal-to-noise level, and it must be close enough so that 
the proper motion exhibited over the life-time of an experiment is above the threshold angular resolution 
of the detector. 
The details of this potential observation were studied in \cite{Koushiappas:2006qq} under the assumption 
that a certain fraction of the Milky Way mass at the solar radius is in microhalos which were all 
formed in a particular epoch. 

Any potential detection of proper motion implies the presence of a large number of ``unresolved'' 
microhalos at distances far beyond the flux-limited distance set by the sensitivity of a particular detector. 
The presence of a large number of microhalos below individual detection threshold would lead to a background 
radiation that would correlate with the angular and radial distribution of microhalos in the Milky Way. 
This was the subject of study in \cite{Pieri:2007ir}, where it was shown that the measured value of 
the $\gamma$-ray background measured by EGRET \cite{Sreekumar:1997un} 
is already placing stringent limits on the probability of 
proper motion detection (due to the implied low number density of microhalos in the Milky Way). 
Addressing the same issue, \cite{Ando:2008br} showed that regardless of the assumptions of the inner 
density profile of microhalos, the EGRET  measurement of the $\gamma$-ray flux from the Galactic center 
\cite{EGRETwebsite} (see \cite{Pullen:2006sy}, but also \cite{Mack:2008wu}) 
is strongly suggesting that the probability of detecting microhalos with measurable proper motion is 
negligible. Nevertheless, given the simplicity of this measurement, 
a search for the proper motion of $\gamma$-ray-only 
sources should be performed due to the valuable information contained in such a potential detection. 

\section{Microhalos and the local Milky Way dark matter distribution }
\label{sec:direct}

The presence of substructure in dark matter halos leads to interesting consequences when it comes to 
direct detection experiments. In general, the rate of events in a direct detection experiment is proportional 
to $\Gamma \sim \rho_\odot$, where $\rho_\odot$ is the dark matter density in the solar neighborhood. 
Typically, the value of $\rho_\odot$ is obtained from dynamical measurements of the structure of the Milky Way 
(e.g. \cite{Klypin:2001xu}), with a canonical value of $\rho_\odot = 0.4 \, {\rm GeV}\, {\rm cm}^{-3}$.
This value 
is obtained from averaging regions which are of order $\sim$kpc, much larger than the regions probed in the 
duration of a direct detection experiment (sub-parsec scales), and it is susceptible to uncertainties that 
are introduced due to the unknown shape of the Milky Way halo (see e.g. \cite{Kamionkowski:1997xg,Zemp:2008gw}). 
If there are fluctuations on sub-parsec scales (as what 
one might expect from the presence of microhalos), then direct detection experiments are prone to 
structure formation uncertainties. 
In this section I give an overview of the experimental implications of 
fluctuations in the density field at the solar radius. 

\subsection{Microhalos and the mean dark matter density}

It is expected that if a certain amount of mass in a dark matter halo is distributed in high-density 
regions, then the density of the smoothly-distributed dark matter will be lower than the case where all 
matter is distributed smoothly. The presence of sub-solar mass microhalos in the Milky Way raises one important 
question: what is the value of the smooth component if the substructure mass function extends down 
to microhalo scales, and what is the probability of the solar system being in an overdense region at 
any given epoch. The answer to both of these questions is directly related to the local flux enhancement in the 
annihilation signal (the so-called ``boost factor'''), as the probability of being in a density enhanced 
region is inversely proportional to the boost factor.

\begin{figure}[!t]
\centerline{
\includegraphics[width=2.5in]{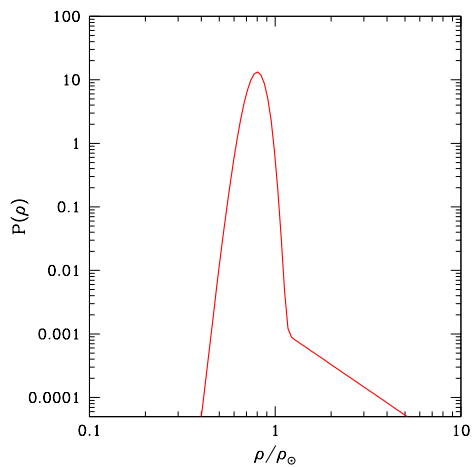}}
\caption{The local dark matter density probability distribution function \cite{Kamionkowski:2008vw}. The peak 
of the distribution defines the dark matter density of the smooth distribution of dark matter, while the power 
low behaviour at higher densities arises from the presence of dark matter subhalos. This function is derived 
under the assumption that the substructure mass fraction in the Milky Way of all substructure in the mass 
range $[10^{-10}M_\odot - M_{\rm MW}]$ is $\sim 20 \%$. }
\label{fig:figure6}
\end{figure}

Addressing this question was the study presented in \cite{Kamionkowski:2008vw}. Under the assumption of 
hierarchical structure formation, it was shown that the derived local density 
probability distribution function (PDF) is positively skewed due to the presence of substructure, and that 
the peak of the distribution is always less than the canonical mean value of the local dark matter density in the 
solar neighborhood (see Fig.~\ref{fig:figure6}). 
This conclusion is derived by a simple analytic model based on the scale-invariant nature 
of the hierarchical structure formation, as well as by investigating the implications of the distribution and 
profiles of dark matter halos as seen in numerical simulations. Both approaches showed similar results. 
These results were recently confirmed in numerical simulations 
by analyzing a suite of halos taken from the Aquarius Project 
\cite{Vogelsberger:2008qb,Springel:2008cc} 
and independendly, by an analysis of the Via Lactea II
Milky Way simulation \cite{Zemp:2008gw,Diemand:2008in}. 

An interesting outcome of the approach of  \cite{Kamionkowski:2008vw} is a derivation of the 
the local annihilation rate due to the granularity of the local dark matter halo. 
The {\it local} boost  factor should not be confused with the {\it global} boost factor that is commonly 
referred in the litereature (see e.g. \cite{Strigari:2006rd,Martinez:2009jh}). 
The least granular cases studied in \cite{Kamionkowski:2008vw} yield a boost factor which is roughly 
consistent with estimates based on the survival rate of microhalos \cite{Berezinsky:2003vn} as well 
as direct simulation results, as shown in \cite{Diemand:2008in}. 
The local boost factor is $\sim [1-5]$, with 
a weak dependence on the cutoff scale of the subhalo mass function.

The introduced variance in the local dark matter density has significant implications to the interpretation of 
combined results from indirect and 
direct detection experiments, as the expected signal is proportional to different powers of the local particle 
density. The volume probed during a 3-year direct detection experiment is very small ($~\sim {\rm few} 
\times 10^{-4} {\rm pc}$), while dynamical estimates of the local dark matter halo provide averages 
over much larger scales. The uncertainties implied by fluctuations in the local dark matter density due 
to the presence of substructure are manifested as uncertainties in the predicted rates in a dark matter 
detector \cite{Kamionkowski:2008vw}. If substructure in the form of microhalos is abundant in the solar 
neighborhood, then the local 
boost factor will be large (favoring indirect detection searches), 
however the probability of being in an overdense region will be low (disfavoring direct detection experiments).
Fortuitously, present simulation results suggest that these effects (and similarly perhaps the 
uncertainties due to structure in phase space) are small, and a smooth dark matter halo is a safe assumption 
\cite{Vogelsberger:2008qb,Springel:2008cc,Zemp:2008gw,Diemand:2008in}, though the consequences for 
indirect detection searches may be more pronounced (e.g., the explanation of the postitron excess in the 
PAMELA \cite{Adriani:2008zr} and ATIC \cite{:2008zzr} data by the presence of a nearby 
subhalo \cite{Hooper:2008kv}). 

\begin{figure}[!t]
\centerline{\includegraphics[width=2.5in]{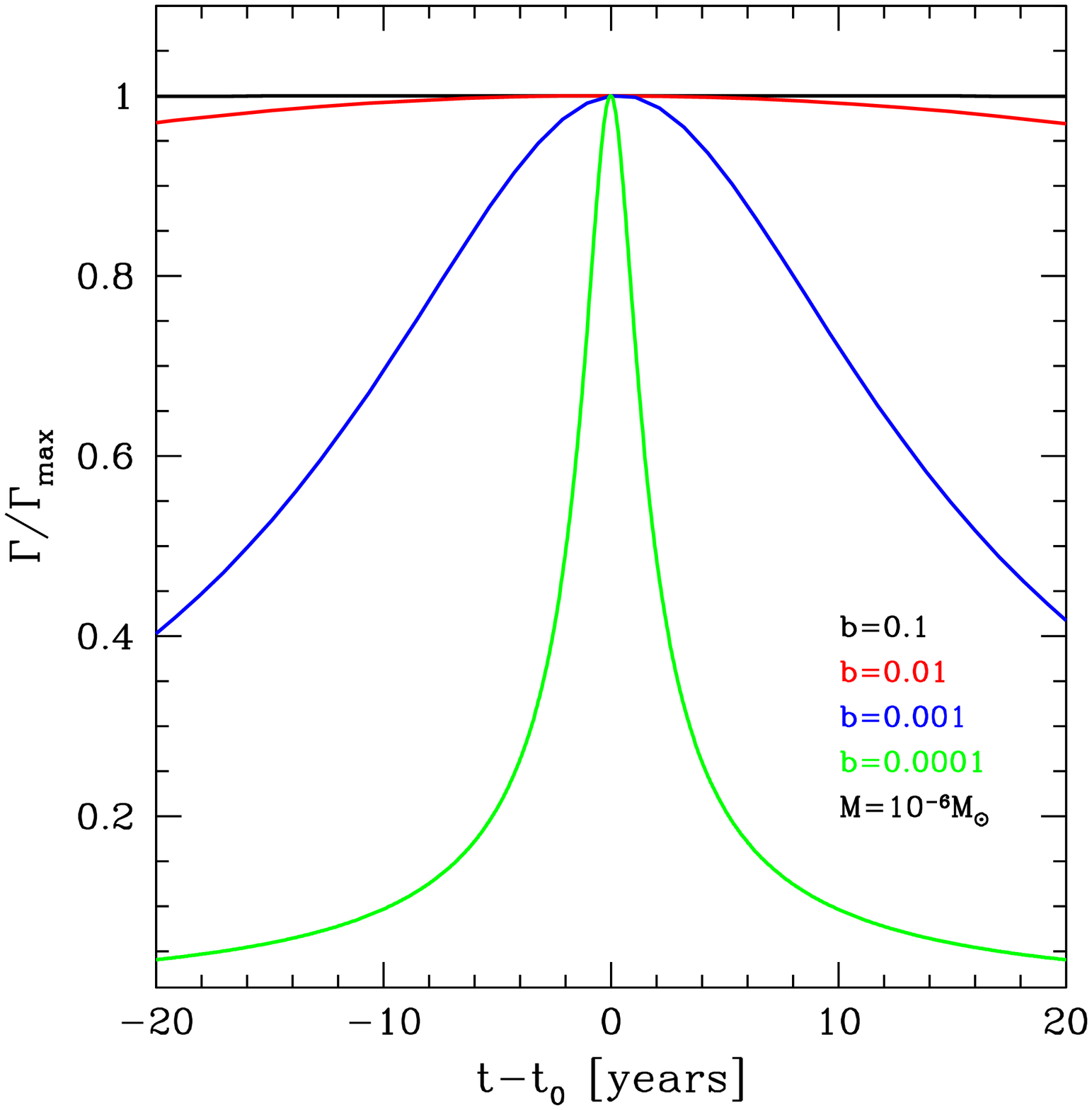}
\includegraphics[width=2.5in]{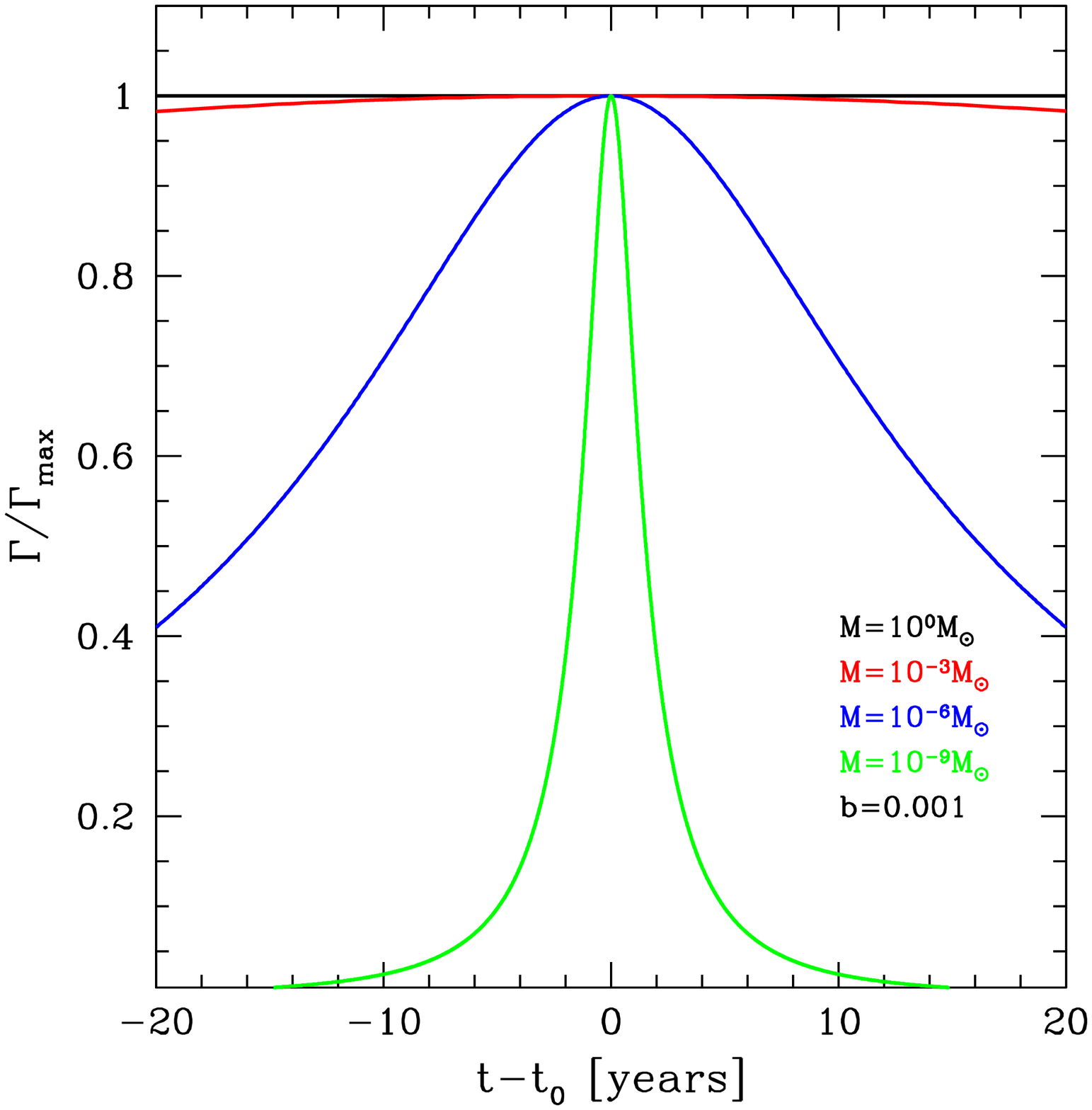}}
\caption{The time-dependent flux enhancement expected from a toy-example of a passage of the solar system through a 
$10^{-6} M_\odot$ microhalo. The {\em left} panel shows the effects of the impact parameter on the flux enhancement 
(normalized to the maximum flux at closest approach), while the {\em right} panel shows the same quantity but at a 
fixed impact parameter, for different mass microhalos.}
\label{fig:figure7}
\end{figure}

An interesting question would be to ask whether it is possible that a long-duration direct detection 
experiment will be able to map the local distribution of dark matter. If we assume the velocity of the 
Sun to be 220 km/s, then the distance sweept over $\sim 20$ years, is 
$x \sim {\rm few } \times 10^{-3} \, {\rm pc}$. 
An extrapolation of the subhalo mass function to microhalo masses (normalized so that 10\% of the Milky Way 
halo is in objects with mass greater than $10^{-5}M_{\rm MW}$) and assuming an NFW profile for the radial 
distribution of subhalos, results in a mean distance between microhalos which depends on mass 
as $\langle r \rangle \approx 52 M^{0.3} \, {\rm pc}$. For microhalos of mass $10^{-6} M_\odot$, this is 
$\langle r \rangle \approx 0.8\, {\rm pc}$. As $x \ll \langle r \rangle$, it seems 
unlikely that a direct detection experiment will measure the transition from a smooth component 
to an overdensity of dark matter. Coupled with the very small likelihood of being inside a microhalo at the 
present time (see \cite{Kamionkowski:2008vw,Vogelsberger:2008qb,Zemp:2008gw}) 
it implies that the rate in direct detection experiments should be 
time independent (note that this statement does not include the annual modulation that should be present due 
to the relative motions of the Sun and the Earth in the Milky Way dark matter halo). 
Fig.~\ref{fig:figure7} shows the expected time dependence of the signal in a long-duration direct detection 
experiment in the hypothetical scenario where the solar system passes through a microhalo. It is important to 
emphasize that this effects are not convolved with the likelihood of such an occurrence, they simply demonstrate 
the magnitude of any effect for each particular case shown.

\section{Conclusion} 
\label{sec:conclusions}
Sub-solar mass dark matter halos are interesting, not only because they are linked to the nature of dark matter, 
but also because a detection of their presence would provide insight into structure formation at extremely early times. 
The detection of microhalos would first and foremost show that the dark matter particle is cold. In addition, 
it will place constraints on the value of the kinetic decoupling temperature and mass of the dark matter particle. 
From the cosmological perspective, any detection of sub-solar mass halos would provide insights into halo merging 
and growth at extremely high redshifts, a task unattainable by any other observational method. Their mere 
presence would imply that at least a certain fraction of them survived the rapid merger phase.
Regardless of the difficulties presented by these highly non-linear structures, 
sub-solar mass dark matter halos are extremely interesting objects and 
certainly warrant further investigations.

\ack
I thank the referee, Juerg Diemand, for numerous comments that improved the quality and content of this manuscript. 
I acknowledge useful conversations with Alex Geringer-Sameth, 
Richard Gaitskell, Jerry Jungman, and 
Marc Kamionkowski. This work was supported by Brown University.


\section*{References}

\bibliography{njp09}

\begin{thebibliography}{10}

\bibitem{Komatsu:2008hk}
E.~Komatsu et~al.
\newblock {Five-Year Wilkinson Microwave Anisotropy Probe
  Observations:Cosmological Interpretation}.
\newblock {\em Astrophys. J. Suppl.}, 180:330--376, 2009.

\bibitem{Jungman:1995df}
Gerard Jungman, Marc Kamionkowski, and Kim Griest.
\newblock Supersymmetric dark matter.
\newblock {\em Phys. Rept.}, 267:195--373, 1996.

\bibitem{BHS05}
Gianfranco Bertone et~al.
\newblock Particle dark matter: Evidence, candidates and constraints.
\newblock {\em Phys. Rept.}, 405:279--390, 2005.

\bibitem{HSS01}
Stefan Hofmann et~al.
\newblock Damping scales of neutralino cold dark matter.
\newblock {\em Phys. Rev.}, D64:083507, 2001.

\bibitem{CKZ02}
Xuelei Chen, Marc Kamionkowski, and Xinmin Zhang.
\newblock Kinetic decoupling of neutralino dark matter.
\newblock {\em Phys. Rev.}, D64(2):021302, Jun 2001.

\bibitem{Profumo:2006bv}
Stefano Profumo, Kris Sigurdson, and Marc Kamionkowski.
\newblock What mass are the smallest protohalos?
\newblock {\em Phys. Rev. Lett.}, 97:031301, 2006.

\bibitem{Schmid:1998mx}
Christoph Schmid et~al.
\newblock Amplification of cosmological inhomogeneities from the qcd
  transition.
\newblock {\em Phys. Rev.}, D59:043517, 1999.

\bibitem{Green:2003un}
Anne~M. Green, Stefan Hofmann, and Dominik~J. Schwarz.
\newblock The power spectrum of susy-cdm on sub-galactic scales.
\newblock {\em Mon. Not. Roy. Astron. Soc.}, 353:L23, 2004.

\bibitem{Green:2005fa}
Anne~M. Green et~al.
\newblock The first wimpy halos.
\newblock {\em JCAP}, 0508:003, 2005.

\bibitem{LZ05}
Abraham Loeb and Matias Zaldarriaga.
\newblock The small-scale power spectrum of cold dark matter.
\newblock {\em Phys. Rev.}, D71:103520, 2005.

\bibitem{Martinez:2009jh}
Gregory~D. Martinez, James~S. Bullock, Manoj Kaplinghat, Louis~E. Strigari, and
  Roberto Trotta.
\newblock {Indirect Dark Matter Detection from Dwarf Satellites: Joint
  Expectations from Astrophysics and Supersymmetry}.
\newblock 2009.

\bibitem{DMS05}
Jurg Diemand et~al.
\newblock Earth-mass dark-matter haloe s as the first structures in the early
  universe.
\newblock {\em Nature.}, 433:389--391, 2005.

\bibitem{Bertschinger:2001ng}
Edmund Bertschinger.
\newblock {Multiscale Gaussian Random Fields for Cosmological Simulations}.
\newblock {\em Astrophys. J. Suppl.}, 137:1, 2001.

\bibitem{Diemand:2006ey}
Jurg Diemand, Michael Kuhlen, and Piero Madau.
\newblock Early supersymmetric cold dark matter substructure.
\newblock {\em Astrophys. J.}, 649:1--13, 2006.

\bibitem{Elahi:2008cd}
P.~J. Elahi, R.~J. Thacker, L.~M. Widrow, and E.~Scannapieco.
\newblock {Subhaloes in Scale-Free Cosmologies}.
\newblock 2008.

\bibitem{Bode:2000gq}
Paul Bode, Jeremiah~P. Ostriker, and Neil Turok.
\newblock {Halo Formation in Warm Dark Matter Models}.
\newblock {\em Astrophys. J.}, 556:93--107, 2001.

\bibitem{Knebe:2003hs}
Alexander Knebe, Julien E.~G. Devriendt, Brad~K. Gibson, and Joseph Silk.
\newblock {Top-Down Fragmentation of a Warm Dark Matter Filament}.
\newblock {\em Mon. Not. Roy. Astron. Soc.}, 345:1285, 2003.

\bibitem{Yoshida:2003rm}
Naoki Yoshida, Aaron Sokasian, Lars Hernquist, and Volker Springel.
\newblock {Early Structure Formation and Reionization in a Warm Dark Matter
  Cosmology}.
\newblock {\em Astrophys. J.}, 591:L1--L4, 2003.

\bibitem{Colin:2000dn}
Pedro Colin, Vladimir Avila-Reese, and Octavio Valenzuela.
\newblock {Substructure and halo density profiles in a Warm Dark Matter
  Cosmology}.
\newblock {\em Astrophys. J.}, 542:622--630, 2000.

\bibitem{Goetz:2002vm}
Martin Goetz and Jesper Sommer-Larsen.
\newblock {Galaxy Formation: Warm Dark Matter, Missing Satellites, and the
  Angular Momentum Problem}.
\newblock {\em Astrophys. Space Sci.}, 284:341--344, 2003.

\bibitem{Heitmann:2006eu}
Katrin Heitmann et~al.
\newblock Capturing halos at high redshifts.
\newblock {\em Astrophys. J.}, 642:L85--L88, 2006.

\bibitem{1996clss.conf..349W}
S.~D.~M. {White}.
\newblock {Formation and Evolution of Galaxies}.
\newblock In R.~{Schaeffer}, J.~{Silk}, M.~{Spiro}, and J.~{Zinn-Justin},
  editors, {\em Cosmology and Large Scale Structure}, pages 349--+, January
  1996.

\bibitem{Berezinsky:2003vn}
Veniamin Berezinsky et~al.
\newblock Small-scale clumps in the galactic halo and dark matter annihilation.
\newblock {\em Phys. Rev.}, D68:103003, 2003.

\bibitem{ZSH1}
HongSheng Zhao et~al.
\newblock Earth-mass dark matter halos are torn into dark mini-streams by
  stars.
\newblock 2005.

\bibitem{METAL05}
Ben Moore et~al.
\newblock On the survival and disruption of earth mass cdm micro- haloes.
\newblock 2005.

\bibitem{Green:2006hh}
Anne~M. Green and Simon~P. Goodwin.
\newblock Mini-halo disruption due to encounters with stars.
\newblock {\em Mon. Not. Roy. Astron. Soc.}, 375:1111--1120, 2007.

\bibitem{Goodwin:2003fn}
Simon~P. Goodwin, A.~P. Whitworth, and D.~Ward-Thompson.
\newblock {Simulating star formation in molecular cloud cores I. The influence
  of low levels of turbulence on fragmentation and multiplicity}.
\newblock {\em Astron. Astrophys.}, 414:633--650, 2004.

\bibitem{Goodwin:2004yb}
Simon~P. Goodwin, Anthony~P Whitworth, and Derek Ward-Thompson.
\newblock {Simulating star formation in molecular cores II. The effects of
  different levels of turbulence}.
\newblock {\em Astron. Astrophys.}, 423:169--182, 2004.

\bibitem{Goerdt:2006hp}
Tobias Goerdt et~al.
\newblock The survival and disruption of cdm micro-haloes: implications for
  direct and indirect detection experiments.
\newblock {\em Mon. Not. Roy. Astron. Soc.}, 375:191--198, 2007.

\bibitem{NFW96}
Julio~F. Navarro et~al.
\newblock The structure of cold dark matter halos.
\newblock {\em Astrophys. J.}, 462:563--575, 1996.

\bibitem{DiemandPC}
J{\"{u}}rg Diemand.
\newblock Private Communication, 2009.

\bibitem{Bryan:1997dn}
G.~L. Bryan and M.~L. Norman.
\newblock {Statistical Properties of X-ray Clusters: Analytic and Numerical
  Comparisons}.
\newblock {\em Astrophys. J.}, 495:80, 1998.

\bibitem{Bullock:1999he}
James~S. Bullock et~al.
\newblock Profiles of dark haloes: evolution, scatter, and environment.
\newblock {\em Mon. Not. Roy. Astron. Soc.}, 321:559--575, 2001.

\bibitem{Diemand:2005rd}
Jurg Diemand, Piero Madau, and Ben Moore.
\newblock {The distribution and kinematics of early high-sigma peaks in
  present-day haloes: implications for rare objects and old stellar
  populations}.
\newblock {\em Mon. Not. Roy. Astron. Soc.}, 364:367--383, 2005.

\bibitem{Klypin:2001xu}
Anatoly Klypin, HongSheng Zhao, and Rachel~S. Somerville.
\newblock {LCDM-based models for the Milky Way and M31 I: Dynamical Models}.
\newblock {\em Astrophys. J.}, 573:597--613, 2002.

\bibitem{Eke:2000av}
Vincent~R. Eke, J.~F. Navarro, and M.~Steinmetz.
\newblock {The Power Spectrum Dependence of Dark Matter Halo Concentrations}.
\newblock {\em Astrophys. J.}, 554:114--125, 2001.

\bibitem{2005AAS...207.2405R}
S.~{Ritz}, J.~{Grindlay}, C.~{Meegan}, P.~F. {Michelson}, and {GLAST Mission
  Team}.
\newblock {The Gamma-ray Large Area Space Telescope (GLAST) Mission}.
\newblock In {\em Bulletin of the American Astronomical Society}, pages
  1198--+, December 2005.

\bibitem{Weekes:2001pd}
T.~C. Weekes et~al.
\newblock Veritas: The very energetic radiation imaging telescope array system.
\newblock {\em Astropart. Phys.}, 17:221--243, 2002.

\bibitem{2003ICRC....5.2811H}
W.~{Hofmann} and {H.~E.~S.~S.~Collaboration}.
\newblock {Status of the H.E.S.S. Project}.
\newblock In {\em International Cosmic Ray Conference}, page 2811, July 2003.

\bibitem{2003ICRC....5.2815M}
M.~{Martinez} and {MAGIC Collaboration}.
\newblock {Status of the MAGIC Telescope}.
\newblock In {\em International Cosmic Ray Conference}, page 2815, July 2003.

\bibitem{Yoshikoshi:1999rg}
T.~Yoshikoshi et~al.
\newblock Present status of the 7-m to 10-m telescope of cangaroo ii.
\newblock {\em Astropart. Phys.}, 11:267--269, 1999.

\bibitem{KZW04}
S.~Koushiappas et~al.
\newblock {Observability of gamma rays from neutralino annihilations in the
  Milky Way substructure}.
\newblock {\em Phys. Rev.}, 69(4):043501--+, February 2004.

\bibitem{Kuhlen:2008aw}
Michael Kuhlen, Jurg Diemand, and Piero Madau.
\newblock {The Dark Matter Annihilation Signal from Galactic Substructure:
  Predictions for GLAST}.
\newblock 2008.

\bibitem{Sommerfeld:1931}
A.~Sommerfeld.
\newblock {\em Ann. der Physik}, 403:257, 1931.

\bibitem{Hisano:2003ec}
Junji Hisano, Shigeki Matsumoto, and Mihoko~M. Nojiri.
\newblock {Explosive dark matter annihilation}.
\newblock {\em Phys. Rev. Lett.}, 92:031303, 2004.

\bibitem{Profumo:2005xd}
Stefano Profumo.
\newblock {TeV gamma-rays and the largest masses and annihilation cross
  sections of neutralino dark matter}.
\newblock {\em Phys. Rev.}, D72:103521, 2005.

\bibitem{ArkaniHamed:2008qn}
Nima Arkani-Hamed, Douglas~P. Finkbeiner, Tracy~R. Slatyer, and Neal Weiner.
\newblock {A Theory of Dark Matter}.
\newblock {\em Phys. Rev.}, D79:015014, 2009.

\bibitem{MarchRussell:2008tu}
John~David March-Russell and Stephen~Mathew West.
\newblock {WIMPonium and Boost Factors for Indirect Dark Matter Detection}.
\newblock 2008.

\bibitem{Pieri:2005pg}
Lidia Pieri et~al.
\newblock Difficulty of detecting minihalos via gamm rays from dark matter
  annihilation.
\newblock {\em Phys. Rev. Lett.}, 95:211301, 2005.

\bibitem{Koushiappas:2006qq}
Savvas~M. Koushiappas.
\newblock {Proper motion of gamma-rays from microhalo sources}.
\newblock {\em Phys. Rev. Lett.}, 97:191301, 2006.

\bibitem{Koushiappas:2007mx}
Savvas~M. Koushiappas.
\newblock {Detecting the dark matter via the proper motion of gamma- rays from
  microhalos}.
\newblock {\em AIP Conf. Proc.}, 921:142--146, 2007.

\bibitem{Pieri:2007ir}
L.~Pieri, G.~Bertone, and E~Branchini.
\newblock {Dark Matter Annihilation in Substructures Revised}.
\newblock {\em Mon. Not. Roy. Astron. Soc.}, 384:1627, 2008.

\bibitem{Sreekumar:1997un}
P.~Sreekumar et~al.
\newblock Egret observations of the extragalactic gamma ray emission.
\newblock {\em Astrophys. J.}, 494:523--534, 1998.

\bibitem{Ando:2008br}
Shin'ichiro Ando, Marc Kamionkowski, Samuel~K. Lee, and Savvas~M. Koushiappas.
\newblock {Can proper motions of dark-matter subhalos be detected?}
\newblock {\em Phys. Rev.}, D78:101301, 2008.

\bibitem{EGRETwebsite}
http://heasarc.gsfc.nasa.gov/docs/cgro/egret.

\bibitem{Pullen:2006sy}
Anthony~R. Pullen, Ranga-Ram Chary, and Marc Kamionkowski.
\newblock {Search with EGRET for a gamma ray line from the galactic center}.
\newblock {\em Phys. Rev.}, D76:063006, 2007.

\bibitem{Mack:2008wu}
Gregory~D. Mack, Thomas~D. Jacques, John~F. Beacom, Nicole~F. Bell, and Hasan
  Yuksel.
\newblock {Conservative Constraints on Dark Matter Annihilation into Gamma
  Rays}.
\newblock {\em Phys. Rev.}, D78:063542, 2008.

\bibitem{Kamionkowski:1997xg}
Marc Kamionkowski and Ali Kinkhabwala.
\newblock {Galactic halo models and particle dark matter detection}.
\newblock {\em Phys. Rev.}, D57:3256--3263, 1998.

\bibitem{Zemp:2008gw}
Marcel Zemp et~al.
\newblock {The Graininess of Dark Matter Haloes}.
\newblock 2008.

\bibitem{Kamionkowski:2008vw}
Marc Kamionkowski and Savvas~M. Koushiappas.
\newblock {Galactic Substructure and Direct Detection of Dark Matter}.
\newblock {\em Phys. Rev.}, D77:103509, 2008.

\bibitem{Vogelsberger:2008qb}
Mark Vogelsberger et~al.
\newblock {Phase-space structure in the local dark matter distribution and its
  signature in direct detection experiments}.
\newblock 2008.

\bibitem{Springel:2008cc}
Volker Springel et~al.
\newblock {The Aquarius Project: the subhalos of galactic halos}.
\newblock 2008.

\bibitem{Diemand:2008in}
J.~Diemand et~al.
\newblock {Clumps and streams in the local dark matter distribution}.
\newblock {\em Nature.}, 454:735--738, 2008.

\bibitem{Strigari:2006rd}
Louis~E. Strigari, Savvas~M. Koushiappas, James~S. Bullock, and Manoj
  Kaplinghat.
\newblock {Precise constraints on the dark matter content of Milky Way dwarf
  galaxies for gamma-ray experiments}.
\newblock {\em Phys. Rev.}, D75:083526, 2007.

\bibitem{Adriani:2008zr}
Oscar Adriani et~al.
\newblock {An anomalous positron abundance in cosmic rays with energies 1.5.100
  GeV}.
\newblock {\em Nature}, 458:607--609, 2009.

\bibitem{:2008zzr}
J.~Chang et~al.
\newblock {An excess of cosmic ray electrons at energies of 300.800 GeV}.
\newblock {\em Nature}, 456:362--365, 2008.

\bibitem{Hooper:2008kv}
Dan Hooper, Albert Stebbins, and Kathryn~M. Zurek.
\newblock {The PAMELA and ATIC Excesses From a Nearby Clump of Neutralino Dark
  Matter}.
\newblock 2008.

\end{thebibliography}

\end{document}